\documentclass{elsart}
\usepackage{epsfig}
\begin{document}
\begin{frontmatter}
\title{Implementation of the Dead-Time Free F1 TDC in the COMPASS 
Detector Readout}
\author{H.~Fischer},
\author{J.~Franz},
\author{A.~Gr\"unemaier},
\author{F.H.~Heinsius\thanksref{FHH}},
\author{L.~Hennig},
\author{K.~K\"onigsmann},
\author{M.~Niebuhr},
\author{T.~Schmidt},
\author{H.~Schmitt},
\author{H.J.~Urban}
\address{Fakult\"at f\"ur Physik, Universit\"at Freiburg, 
79104 Freiburg, Germany}
\thanks[FHH]{Corresponding author; e-mail: heinsius@cern.ch.}

\begin{abstract}
In the scope of the COMPASS experiment a multi-purpose, high rate
capable TDC with digitisation width of 60~ps has been developed. 
The integration into the readout system and the flexible input
of the CATCH readout driver is presented.
\end{abstract}
\end{frontmatter}

The COMPASS experiment at CERN will investigate the hadron structure 
by deep inelastic muon scattering and hadronic production 
processes.
One central issue of the experimental effort will be the measurement of
the contribution of gluons to the nucleon spin. 
To reach this objective a demanding state-of-the-art double-stage
spectrometer with large geometrical and dynamical acceptance is being
set-up and commissioned through the year 2000.

\section{Overview of the readout architecture}

\begin{figure}[htb]
\epsfig{file=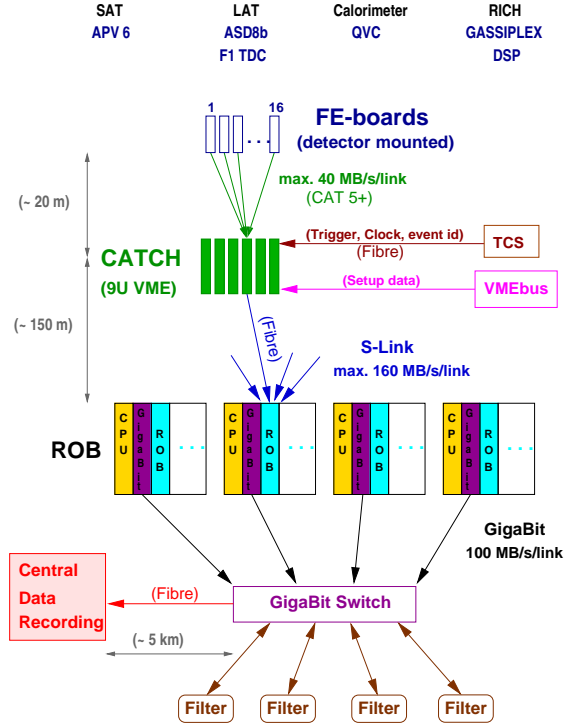, width=7.5cm}
\caption{The architecture of the COMPASS detector readout.}
\label{fig:architecture}
\end{figure}

The large active detector volume, the
necessary expandability to perform different physics experiments and 
the upgradeability during the longevity of the experiment  
requires a scalable and distributed readout system with federal 
event building (Figure~\ref{fig:architecture}). 
The high particle flux from the traversing beam at
intensities of $2\cdot 10^8$\
particles per spill (2.4~s) requires
dead-timeless readout with digitisation
immediately at the detector. Data are stored on the front-end boards 
in random access memories
or pipelines for about 2~$\mu$s until trigger decisions have been taken. 
Only those hits are transmitted to the master event-builder which are 
time correlated to a physics trigger event. 
Hence background data are suppressed at an early stage of the readout chain,
which reduces the necessary bandwidth significantly down to several
Gigabytes per spill.

The trigger rate depends on the physics 
reactions to be studied and is in the range of 10--100~kHz. 
Trigger signals are distributed synchronously to
a 38.88~MHz clock via optical fibres from the trigger control system
(TCS) laser crate. Additional event information like spill and event numbers,
time synchronisation signal and begin- and end-of-spill flags are
transmitted through the same fibre by time multiplexing.

The readout driver named CATCH (COMPASS Accumulate, Transfer and
Control Hardware~\cite{leb98}) serves as a common interface between the
detector-specific front-end boards,
the trigger control system and the readout buffers.
Data which belong to one event are combined in a local sub-event building 
process and consistency checks are performed on the data integrity.
The sub-events are transmitted via a fast optical link to readout buffers,
which can store all data from at least one spill. Sub-events are combined
and  transmitted via
Gigabit Ethernet to filter computers. Here the
final event-building is performed, events are partly reconstructed and,
based on physics cuts, data may be reduced by a factor of 5
to 10. Finally a continuous rate of 30~MByte/s of event data 
can be transfered to the central data recording facilities at CERN, 
where the events will be stored in an Objectivity/DB data base. 
The total amount of data accumulated each year will be about 300~Terabyte.

\section{F1-TDC}

To be in compliance with a dead-time free data-digitisation  
and the time resolution requirements of the detectors 
we developed a new TDC-chip.
This device can be easily adapted
to the conditions of different detector types~\cite{leb99}.
The F1-TDC, produced in 0.6~$\mu$m sea-of-gates
technology, supports basically three modes:
In the high rate option ($>$~6~MHz input rate) the TDC chip provides a
digitisation width of 60~ps on four input channels at a double pulse
resolution of about 18~ns.
The standard mode delivers a digitisation unit of about 120~ps on
eight inputs per chip at a dynamic range of 16 bits.
The latch mode implements a
32~channel low resolution ($>$4.7~ns) readout for multi wire
proportional chambers.
In all modes the hits, which are time correlated to the trigger.
are selected for readout from an internal buffer.

We have chosen two options for interfacing the F1-TDC to the CATCH
readout driver:
For the high rate detectors like scintillating fibres, trigger
hodoscopes and the beam momentum stations, the TDCs are placed on
mezzanine cards, which are mounted on the CATCH, allowing for a
readout speed of 120~MByte/s.
Each mezzanine card hosts four TDC chips and supports 32 differential
inputs in
standard resolution or 16~inputs in high resolution mode (software
selectable). 
The straw tubes, drift chambers and micromega chambers have the TDCs
mounted on the front-end boards operated in standard resolution mode,
as well as the multi wire proportional chambers with the F1-chip in
latch mode.
In case of the micromega chambers~\cite{micromega} both leading and
trailing edges are read-out allowing for noise suppression and walk 
correction by measuring the time over threshold.
For all front-end mounted F1-TDCs a DAC interface is used to set the
thresholds of the external discriminators.
Up to eight TDCs are placed on one front-end board and read-out
through a common bus, which interfaces to a HOTLink
serialiser chip. 

\section{CATCH readout driver}

\begin{figure}[htb]
\epsfig{file=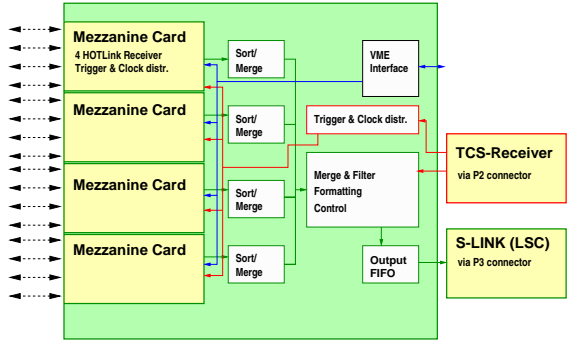, width=7.5cm}
\caption{CATCH schematic.}
\label{fig:catch}
\end{figure}

Four exchangeable input mezzanine cards
mounted on the 9U VME module provide a highly flexible input interface
(Figure~\ref{fig:catch}). 
At the moment we have developed three different types of
mezzanine cards following the IEEE CMC standard:

One type receives digitised data from up to four different 
front-end boards. For this mezzanine card we use the HOTLink 
serialising chip and a 600~MHz differential twisted pair line to 
achieve 400 MBaud sustained input data rate.
The same patch cable is used to transmit trigger, time
synchronisation and a power-up reset signal on a common line.
Through a third and fourth line the reference clock 
and initialisation data, respectively, are distributed to the 
front-end boards. This card is also available with a 
fibre optical link to replace copper cables.

Another type of mezzanine card hosts four  F1-TDC chips resulting in a 128
channel TDC module. 
A 32 channel 300 MHz dead-time free and trigger-latency
correcting scaler is implemented on the third mezzanine card. 

While framed data are transmitted at a maximum throughput of 160~MByte/s
to the readout buffer via a S-LINK module~\cite{slink}, an
independent spy buffer can be used to access
events via the VME bus at a lower data rate. This allows for an easy setup
of a stand alone DAQ without S-LINK and spill buffer at test beams.

The live insertion of the CATCH and plug~\& play feature for front-end
and mezzanine boards greatly enhances the maintainability of the system
and eliminates miscabling.
The front-end boards transmit their identification and position on the
detector to the CATCH, which stores this information in the geometry
database and receives the corresponding initialisation data from the
calibration database.

\begin{ack}
This work would be impossible without the formidable support of the local
electronic workshop and engineer team.   
We deeply appreciate the many
stimulating discussions with our colleagues from the COMPASS
collaboration involved in front-end
electronics development. The developments described in this report are
supported by the German Bundesministerium f\"ur Bildung und Forschung.
\end{ack}

\end{document}